\documentclass[12pt]{article}
\usepackage{amsmath}
\usepackage{xcolor}
\usepackage{amssymb}
\usepackage{graphicx}
\usepackage{caption}
\usepackage{float}
\usepackage{bm}
\date{}
\begin{document}

\title{Structure of Polytropic Stars in General Relativity}

\author{Mayer Humi\thanks {e-mail: mhumi@wpi.edu.}\; and John Roumas\thanks{e-mail:roumas@verizon.net}\\
Department of Mathematical Sciences, \\
Worcester Polytechnic Institute, \\
Worcester, MA 01609}

\maketitle

\begin{abstract}

The inner structure of a star or a primordial interstellar cloud is a major 
topic in classical and relativistic physics. The impact that General 
Relativistic principles have on this structure has been the subject of 
many research papers. In this paper we consider within the context of 
General Relativity a prototype model for this problem by assuming that 
a star consists of polytropic gas. To justify this assumption we observe 
that stars undergo thermodynamically irreversible processes and emit heat 
and radiation to their surroundings. Due to the emission of this energy it is
worthwhile to consider an idealized model in which the gas is 
polytropic. To find interior solutions to the Einstein equations of General 
Relativity in this setting we derive a single equation for the cumulative mass 
distribution of the star and use Tolman-Oppenheimer-Volkoff equation to 
derive formulas for the isentropic index and coefficient.
Using these formulas we present analytic and numerical solutions for the 
polytropic structure of self-gravitating stars and examine their stability. 
We prove also that when the thermodynamics of a star as represented by the 
isentropic index and coefficient is known, the corresponding matter density 
within the star is uniquely determined.

\end{abstract}

\thispagestyle{empty}

\newpage
\section{Introduction}

Mass density pattern within a star is an important problem
and has been the subject of intense ongoing research.
Within the context of classical physics Euler-Poisson equations
form the basis for this research \cite{HR50,HR50a}. A special set of  
solutions to these equations for non-rotating spherically symmetric 
stars with mass-density $\rho=\rho(r)$ and flow field $\bf u=0$ is 
provided by the Lane-Emden functions. The generalization of these equations 
to include axi-symmetric rotations was by considered by Milne\cite{HR1}, 
Chandrasekhar\cite{HR1a,HR1b} and many others. 

Another aspect of this problem relates to the emergence of density pattern 
within a primordial interstellar gas. This problem was considered 
first by Laplace in 1796 who conjectured that a primitive interstellar gas 
cloud may evolve under the influence of gravity to form a system of isolated 
rings which may in turn lead to the formation of planetary systems
\cite{HR10a,HR10b,HR10c}. Such a system of rings around a protostar has been 
observed recently in the constellation Taurus\cite{HR21}.

It is obvious however that on physical grounds this problem should treated 
within the context of General relativity. The Einstein equations of 
General Relativity are highly nonlinear \cite{HR1,HR2} and their solution 
presents a challenge that has been addressed by many researchers 
\cite{HR2,HR3,HR3a}. An early solution of these equations 
is due to Schwarzschild for the field exterior to a spherical star \cite{HR4}. 
However, interior solutions (inside space occupied by matter) are especially 
difficult due to the fact that the energy-momentum tensor is not zero. 
Static solutions for this case were derived under idealized assumptions
(such as constant density) by Tolman\cite{HR7}, Adler\cite{HR10,HR2}, 
Buchdahl\cite{HR9} and were addressed more recently in the lecture series by 
Gourgoulhon\cite{HR19} and the review by Paschalidis and Stergioulas
\cite{HR20,HR3,HR4,HR5,HR6,HR10c} (these references contain a 
a lengthy list of publications on this topic).
In addition various constraints were derived for the structure of a 
spherically symmetric body in static gravitational equilibrium 
\cite{HR7,HR8,HR9,HR10,HR10a}. Interior solutions in the presence of 
anisotropy and other geometries were considered also 
\cite{HR11,HR11a,HR12,HR13}. An exhaustive list of references for 
exact solutions of the Einstein equations appears in \cite{HR2,HR3}.

Due to the physical complexity of star interiors which involves several 
concurrent physical processes we consider in this paper an idealized model based on General Relativity in which the star (or the interstellar gas cloud) 
is polytropic and inquire about the mass density pattern within the star 
under this assumption. This model takes into account 
some of the thermodynamic processes within a star which have been ignored 
so far in the literature. To justify the imposed idealizations we observe that 
stars undergo thermodynamically irreversible processes and emit heat and 
radiation to their surroundings. Due the emission of this energy one can 
envision a situation in which the gas entropy within a star remains nearly 
constant.

For polytropic gas we have the following relationship between
pressure $p$ and density $\rho$
\begin{equation}
\label{1.1}
p=A\rho^{\alpha}
\end{equation}
where $\alpha$ is the {\bf isentropy index} and $A$ is the 
{\bf isentropy coefficient}. 
In the literature when $\alpha=1$ the gas is considered to be isothermal. 
However, when (and only when) $\alpha$ equals the ratio of specific heat at constant 
pressure or specific heat at constant volume, the gas is isentropic. For all
other values of $\alpha$ the gas is called polytropic. This marks finite
heat exchanges within the fluid. However, one can consider a 
more general functional relationship between $p$ and $\rho$ where both 
$\alpha$ and $A$ are dependent on $r$. In this paper, however, we restrict
ourselves and consider only functional relationships between $p$ and $\rho$ 
in which only one of these parameters is dependent on $r$, viz. either 
$p=A(r)\rho(r)^{\alpha}$ where $\alpha$ is constant or $p=A\rho(r)^{\alpha(r)}$ 
where $A$ is constant. These two position-dependent expressions for 
the isentropy relationship represent different physical properties of the gas.

The plan of the paper is as follows: In Section $2$ we review the basic 
theory and equations that govern mass distribution and the components of 
the metric tensor. In Section $3$ we derive an equation for the 
cumulative mass of the sphere as a function of $r$ and use the 
Tolman-Oppenheimer-Volkoff (TOV) equation to derive equations for the 
isentropy index and coefficient. We then prove that when these two 
parameters are predetermined the mass density within the star cannot be 
chosen arbitrarily. In Section $4$ we address the stability of a given 
mass distribution to small perturbations.In Section $5$ we present exact 
and numerical solutions for polytropic spheres with predetermined mass 
density distribution and determine their isentropy coefficients and 
stability. We summarize with some conclusions in Section $6$.

\setcounter{equation}{0}
\section{Review}

In this section we present a review of the basic theory, following chapter 
$14$ in \cite{HR2}.

The general form of the Einstein equations is
\begin{equation}
\label{2.1}
R_{mn}-\frac{1}{2}g_{mn}R=-\frac{8\pi \kappa}{c^2}T_{mn},\,\,\,m,n=0,1,2,3.
\end{equation}
where $R_{mn}$ and $R$ are respectively the contracted form of the 
Riemann tensor $R_{abcd}$ and the Ricci scalar, 
$$
R_{mn}=R^a_{man},\,\,\, R=R^m_m.
$$
$T_{mn}$ is the matter stress-energy tensor, $\kappa$ is Newton's gravitational 
constant, $c$ is the speed of light in a vacuum and $g_{mn}$ is the 
metric tensor.

The general expression for the stress-energy tensor is 
\begin{equation}
\label{2.2}
T_{mn}=\rho u_mu_n+\frac{p}{c^2}(u_mu_n-g_{mn}),
\end{equation}
where $\rho({\bf x})$ is the proper density of matter and $u_m({\bf x})$ 
is the four vector velocity of the flow.

In the following we shall assume that $\rho=\rho(r)$, $p=p(r)$ and
a metric tensor of the form
\begin{equation}
\label{2.3}
g_{mn}=c^2e^{\nu}dt^2-[e^{\lambda}dr^2+r^2(d\phi^2+\sin^2\phi d\theta^2)].
\end{equation}
where $\lambda=\lambda(r)$, $\nu=\nu(r)$ and $r,\phi,\theta$ are the spherical 
coordinates in 3-space.

When matter is static $u_m=(u_0,0,0,0)$ and $T_{mn}$ takes the following form,
\begin{eqnarray}
\label{2.4}
T_{mn} = \left(\begin{array}{cccc}
\rho e^{\nu} &0&0 &0 \\
0&\frac{p}{c^2}e^{\lambda} &0 &0\\
0 &0 &\frac{p}{c^2}r^2&0 \\
0&0&0&\frac{p}{c^2}r^2\sin^2\phi\\
\end{array}
\right).
\end{eqnarray}
After some algebra \cite{HR2,HR7,HR8} one obtains equations for
$\rho$, $p$, $\lambda$, $\nu$ and $m(r)$ (where $m(r)$ is the total mass
of the sphere up to radius $r$). These are
\begin{equation}
\label{2.5}
\frac{dm}{dr}=Br^2\rho
\end{equation}
\begin{equation}
\label{2.6}
e^{-\lambda}=1-\frac{2m}{r}
\end{equation}
\begin{equation}
\label{2.7}
\frac{e^{\lambda}}{r^2}=\frac{1}{r^2}-
\frac{1}{4}\left[\left(\frac{d\nu}{dr}\right)^2-
\frac{d\nu}{dr}\frac{d\lambda}{dr}\right]
+\frac{1}{2r}\left(\frac{d\nu}{dr}+\frac{d\lambda}{dr}\right)-
\frac{1}{2}\frac{d^2\nu}{dr^2}
\end{equation}
\begin{equation}
\label{2.8}
\frac{C}{c^2}p=\frac{1}{r^2}-e^{-\lambda}\left(\frac{1}{r^2}+
\frac{1}{r}\frac{d\nu}{dr}\right)
\end{equation}
where
$$
C=-\frac{8\pi\kappa}{c^2}, B=\frac{4\pi\kappa}{c^2}.
$$

In addition we have the Tolman-Oppenheimer-Volkoff (TOV) equation
which is a consequence of (\ref{2.5})-(\ref{2.8}):
\begin{equation}
\label{2.9}
\frac{1}{c^2}\frac{dp}{dr}=-\frac{m-Cr^3p/2c^2}{r(r-2m)}
\left(\rho+\frac{p}{c^2}\right).
\end{equation}
In the following we normalize $c$ to $1$; $B$ remains $-\frac{C}{2}$.

Assuming that $m(r)$ is known we can solve (\ref{2.7}) algebraically for 
$\lambda$ and substitute the result in (\ref{2.8}) to derive the following 
equation for $\nu$:
\begin{equation}
\label{2.10}
\frac{1}{2}\frac{d^2\nu}{dr^2}+\frac{1}{4}\left(\frac{d\nu}{dr}\right)^2
-\frac{1}{2}\frac{\left(3m-r\frac{dm}{dr}-r\right)\frac{d\nu}{dr}}
{r(2m-r)}-\frac{3m-r\frac{dm}{dr}}{r^2(2m-r)}=0 .
\end{equation}
Although this is a nonlinear equation it can be linearized by the
substitution
\begin{equation}
\label{2.11}
\frac{d\nu}{dr}=2\frac{\frac{du}{dr}}{u}=\frac{d\ln(u^2)}{dr}
\end{equation}
which leads to
\begin{equation}
\label{2.12}
\frac{d^2 u}{dr^2}
-\frac{\left(3m-r\frac{dm}{dr}-r\right)}{r(2m-r)}\frac{du}{dr}
-\frac{3m-r\frac{dm}{dr}}{r^2(2m-r)}u=0.
\end{equation}
\setcounter{equation}{0}
\section{On the Structure of Isentropic Stars}

In this section we consider Isenropic stars and derive general analytic 
expressions for $m(r)$, $\alpha(r)$ and $A(r)$.

\subsection{General Equation for $\bm{m(r)}$}

Using the equations presented in the previous section one can derive 
a single equation for $m(r)$ for a polytropic star where 
both $A$ and $\alpha$ are functions of $r$:
\begin{equation}
\label{7.1}
p=A(r)\rho^{\alpha(r)}.
\end{equation}
To this end we substitute the isentropy relation (\ref{7.1}) in (\ref{2.8})
to obtain
\begin{equation}
\label{7.2}
\rho^{\alpha(r)}=\frac{c^2}{CA(r)}\left\{\frac{1}{r^2}-e^{-\lambda}\left(\frac{1}{r^2}+
\frac{1}{r}\frac{d\nu}{dr}\right)\right\}.
\end{equation}
Substituting (\ref{2.5}) for $\rho$ in (\ref{7.2}), normalizing 
$c$ to $1$ and using the fact that $C=-2B$ it follows that
\begin{equation}
\label{7.3}
\left(\frac{\frac{dm(r)}{dr}}{Br^2}\right)^{\alpha(r)}
=-\frac{1}{2BA(r)}\left\{\frac{1}{r^2}-e^{-\lambda}\left(\frac{1}{r^2}+
\frac{1}{r}\frac{d\nu}{dr}\right)\right\}.
\end{equation}
Substituting (\ref{2.6}) for $\lambda$ in (\ref{7.3}) and solving the
result for $\frac{d\nu}{dr}$ yields
\begin{equation}
\label{7.4}
\frac{d\nu}{dr}=-2\frac{\left(\frac{\frac{dm(r)}{dr}}{Br^2}\right)^{\alpha(r)}
BA(r)r^3+m(r)}{r(2m(r)-r)} .
\end{equation}
Differentiating this equation to obtain an expression for 
$\frac{d^2\nu}{dr^2}$ and substituting in (\ref{2.10}) leads finally
to the following general equation for $m(r)$:  
\begin{eqnarray}
\label{7.5}
&&-2r^{3-2\alpha(r)}B^{1-\alpha(r)}(2m(r)-r)
\left(\frac{dm(r)}{dr}\right)^{\alpha(r)} \\ \notag
&&\left\{A(r)\alpha(r)\frac{d^2m(r)}{dr^2}+
\frac{dm(r)}{dr}\left[A(r)\ln\left(\frac{\frac{dm(r)}{dr}}{Br^2}\right)
\frac{d\alpha(r)}{dr}+\frac{dA(r)}{dr}\right]\right\}+ \\ \notag
&&2r^{2-2\alpha(r)}B^{1-\alpha(r)}A(r)
\left(\frac{dm(r)}{dr}\right)^{\alpha(r)+1}
\left[r\frac{dm(r)}{dr}+m(r)(1+4\alpha(r))-2r\alpha(r)\right]+\\ \notag
&&2r^{5-4\alpha(r)}B^{2-2\alpha(r)}A(r)^2
\left(\frac{dm(r)}{dr}\right)^{2\alpha(r)+1}+2m(r)
\left(\frac{dm(r)}{dr}\right)^{2}=0 .
\end{eqnarray}

This is a highly nonlinear equation but it simplifies considerably when 
$A(r)$ is a constant or $\alpha(r)$ is an integer. 
A solution of this equation can then be used to compute the metric 
coefficients using (\ref{2.6}) and (\ref{7.4}). With this equation it 
is feasible to investigate the dependence of the mass distribution 
on the parameters $\alpha(r)$ and $A(r)$

In view of the difficulty of obtaining analytic solutions 
for (\ref{7.5}) an alternative strategy should be used to investigate the
structure of polytropic stars. Thus if we start with some analytic form
of $\rho$ then we can use (\ref{2.5}) to compute $m(r)$. With this data
it is straightforward to derive differential equations for $\alpha(r)$
and $A(r)$ using the TOV equation {(\ref{2.9}).    

\subsection{Equation for $\bm{\alpha(r)}$ when $\bm{A(r)}$ is Constant}

If we let $A(r)$ in (\ref{7.1}) be constant and  
substitute $p=A\rho^{\alpha(r)}$ in (\ref{2.9}) we obtain after some algebra
the following equation for $\alpha(r)$.
\begin{eqnarray}
\label{7.60}
&& A r(2m(r)-r)\rho(r)^{\alpha(r)}\ln(\rho(r))\frac{d\alpha}{dr}+
Ar\alpha(r)\rho(r)^{\alpha(r)-1}(2m(r)-r)\frac{d\rho}{dr}- \\ \notag
&&[m(r)+ABr^3\rho(r)^{\alpha(r)}][A\rho(r)^{\alpha(r)}+\rho(r)]=0.
\end{eqnarray}

\subsection{Equation for $\bm{A(r)}$ when $\bm{\alpha(r)}$ is Constant}

Following the same strategy as in the previous subsection we obtain a
differential equation for $A(r)$
\begin{eqnarray}
\label{7.61}
&&r(2m(r)-r)\rho(r)^{\alpha}\frac{dA(r)}{dr}+
\alpha A(r)r(2m(r)-r)\rho(r)^{\alpha-1}\frac{d\rho}{dr}-\\ \notag
&&[m(r)+Br^3 A(r)\rho(r)^{\alpha}][A(r)\rho(r)^{\alpha}+\rho(r)]=0.
\end{eqnarray}

Thus in this setting (where $\rho$ is predetermined) one can use 
(\ref{7.60}) or (\ref{7.61}) to compute $\alpha(r)$ or $A(r)$ by solving 
a first order differential equation. Alternatively, (\ref{7.60}) and (\ref{7.61}) 
can be converted to an equation for $\rho$ by using (\ref{2.5}).
We can then choose a functional 
form for either $\alpha(r)$ (and a constant value for $A$ in (\ref{7.60})) 
or $A(r)$ (and a constant value for $\alpha$ in (\ref{7.61})) to determine 
$\rho$ subject to proper boundary conditions. It follows then under the tenets
of General Relativity the density of a polytropic star cannot be assigned 
arbitrarily. The same follows from (\ref{7.5}) when the functional form
$\alpha(r)$ and $A(r)$ is predetermined.

We give several examples.
%GOT THIS FAR STEP 1
\subsection{Equation for $\bm{\rho}$ when $\bm{A(r)}$ and $\bm{\alpha(r)}$ are Constant}

Solving (\ref{7.61}) algebraically for $m(r)$ and substituting in (\ref{2.5}), 
we obtain after some algebra a rather complicated equation for $\rho(r)$ with $A=A(r)$ and $\alpha$ constant.
Therefore we present only a special case of this equation in which both are constant.

With both $\alpha(r)$ and $A(r)$ constant, equations (\ref{7.60})
and (\ref{7.61}) collapse to the following. For brevity, we 
suppress the dependence of $m(r)$ and $\rho(r)$ on $r$:
\begin{equation}
\label{7.62}
A\alpha r(2m-r)\frac{d\rho}{dr}-ABr^3\rho^2(A\rho^{\alpha-1}+1)-
m\rho(A+\rho^{-\alpha+1})=0.
\end{equation}
Algebraically isolating $m$ and substituting in (\ref{2.5})
we obtain the following equation for $\rho$:
\begin{equation}
\label{7.63}
A_2\frac{d^2\rho}{dr^2}+A_{12}\left(\frac{d\rho}{dr}\right)^2+
A_{11}\frac{d\rho}{dr}+A_0=0
\end{equation}
where
$$
A_2=-A\alpha r^2\rho^{\alpha}(2A^2Br^2\rho^{2\alpha}+
2ABr^2\rho^{\alpha+1}+A\rho^{\alpha}+\rho),
$$
$$
A_{12}=Ar^2\alpha\rho^{\alpha-1}(2A^2B\alpha r^2\rho^{2\alpha}+
2A^2Br^2\rho^{2\alpha}-4ABr^2\alpha\rho^{\alpha+1}+ \\
4ABr^2\rho^{\alpha+1}+2A\alpha\rho^{\alpha}+A\rho^{\alpha}+(2-\alpha)\rho),
$$
$$
A_{11}=A\alpha r\rho^{\alpha}(3A^2Br^2\rho^{2\alpha}+
6ABr^2\rho^{\alpha+1}+3Br^2\rho^2-2A\rho^{\alpha}-2\rho),
$$
$$
A_0=-Br^2\rho(\rho^3+3A^3\rho^{3\alpha}+7A^2\rho^{2\alpha+1}+5A\rho^{\alpha+2}).
$$
In particular, when $\alpha=1$ and $A$ is normalized to $1$ (\ref{7.63})
reduces to 
\begin{equation}
\label{7.64}
r\rho(2Br^2\rho+1)\frac{d^2\rho}{dr^2}-
2r(Br^2\rho+1)\left(\frac{d\rho}{dr}\right)^2+
2\rho(1-3Br^2\rho)\frac{d\rho}{dr}+8Br\rho^3=0.
\end{equation}

A similar equation can be derived from (\ref{7.63}) for $\alpha=2$ with $A=1$.

In Fig. \ref{Figure 1} we present the solutions of these two cases ($\alpha=1$ and $\alpha=2$,
each with $A=1$) by the red and blue dashed lines. The boundary conditions on 
$\rho$ are $\rho(0.001)=1$ and $\rho(0.995)=5\times 10^{-3}$. These boundary 
conditions are needed to avoid numerical singularities at $0$ and $1$.

Similarly if we let $A(r)=Dr$ (where $D$ is a constant) then for 
$\alpha=1,2$ we obtain for $\rho$ in Fig. \ref{Figure 1} the solid magenta and green 
lines, respectively.  

Thus we demonstrate that in the context of general relativity the mass density of
polytropic star with $A=A(r)$ and constant $\alpha$ 
cannot be assigned arbitrarily. 

Using (\ref{7.60}) and following the same steps described above we can 
obtain similar equations to the case where $\alpha=\alpha(r)$ and $A$ 
is constant.

\setcounter{equation}{0}
\section{Stability}

In this section we derive equations that determine the stability of 
polytropic stars using the two models that were discussed in (\ref{7.60})
and (\ref{7.61}). We then apply these results to the star model discussed 
in the previous section. 

To implement this objective we introduce a perturbation to a star with initial
cumulative mass distribution $m_0$:
\begin{equation}
\label{9.1}
m(r)=m_0(r)+\epsilon m_1(r).
\end{equation}
The star will then be considered stable if, for a perturbation with
initial value $m_1(0)\ll 1$, $|m_1(r)|$ remains bounded and $m(r) \ge 0$.
It will be considered unstable otherwise. To derive the equation that $m_1$ satisfies
we consider the two polytropic models separately.

\subsection{$\bm{p=A\rho^{\alpha(r)}}$}

To simplify the presentation we shall assume that $A=1$ and $B=1$.
Substituting (\ref{9.1}) in (\ref{7.5}) and using (\ref{2.5})
we obtain to first order in $\epsilon$ the following differential equation
for $m_1(r)$:
%%%%%%%%%%%%%%%%%%%%%%%%%%%%%%%%%%%%%%%%%%%%%%%%%%%%%%%
\iffalse
\begin{eqnarray}
\label{9.2}
&&\left[-2zz rr(2m_0-r)\alpha\right]\frac{d^2m_1}{dr^2}+ 
\left\{-2rr zz(2m_0-r)(1+ln(\rho))(\alpha+1) \\ \notag
&&-\left[(2(2m_0-r))((r^2\rho)^{-1+\alpha}rr\left(\frac{d(r^2\rho)}{dr^2}-2zz rrr\right)\right]\right\}\alpha^2 + \\ \notag
&&\{2rrr(r^3 zz\rho+(2(r^2\rho)^{2\alpha} rrr-2zz)+5zz m_0)\alpha+ \\ \notag
&&2rrr(r^3 zz\rho+rrr r(r^2\rho)^{2\alpha}+zz m_0+r(r^2\rho)^{\alpha+1})+
4r^2m_0\rho\}\frac{dm_1}{dr} \\ \notag
&&\left[(-4zz rr\frac{d(r^2\rho)}{dr}+8ww r^{2-2\alpha})\alpha-
4rr ww\ln(\rho)\frac{d\alpha}{dr}+2ww r^{2-2\alpha}+2r^4\rho^2\right]m_1
\end{eqnarray}
\fi
%%%%%%%%%%%%%%%%%%%%%%%%%%%%%%%%%%%%%%%%%%%%%
\begin{multline}
\label{9.2}
\left[-2Z S(2m_0-r)\alpha\right]\frac{d^2m_1}{dr^2}+ 
\{-2S Z(2m_0-r)(1+\ln(\rho))(\alpha+1) \\ 
-\left[2(2m_0-r)\left((r^2\rho)^{-1+\alpha}S\frac{d(r^2\rho)}{dr^2}-2Z R\right)\right]\alpha^2 + \\
2R\left(r^3 Z\rho+r(2(r^2\rho)^{2\alpha} R-2Z)+5Z m_0\right)\alpha+ \\ 
2R(r^3 Z\rho+R r(r^2\rho)^{2\alpha}+Z m_0+r(r^2\rho)^{\alpha+1})+
4r^2m_0\rho \}\frac{dm_1}{dr} \\ 
\left[(-4Z S\frac{d(r^2\rho)}{dr}+8W r^{2-2\alpha})\alpha-
4S W\ln(\rho)\frac{d\alpha}{dr}+2W r^{2-2\alpha}+2r^4\rho^2\right]m_1=0
\end{multline}

where
$$
R=r^{2-2\alpha},\,\,S=r^{3-2\alpha},
W=\left(\frac{dm_0}{dr}\right)^{\alpha+1},\,\,
Z=\left(\frac{dm_0}{dr}\right)^{\alpha}.
$$
\subsection{$\bm{p=A(r)\rho^{\alpha}}$}

For simplicity we treat here only the case $\alpha=1$. Following the same
steps as in the previous subsection we obtain
\begin{eqnarray}
\label{9.3}
&&-Ar\rho(2m_0-r)\frac{d^2m_1}{dr^2}+ \\ \notag
&&\left[r(r-2m_0)(2\rho\frac{dA}{dr}+A\frac{d\rho}{dr})+3\rho^2 Ar^3(A(r)+1)+
2\rho A(3m_0-r)+2\rho m_0\right]\frac{dm_1}{dr}+ \\ \notag
&&r^2\rho\left[(-2r\frac{d\rho}{dr}+\rho)A-2r\rho\frac{dA}{dr}+\rho\right]m_1=0
\end{eqnarray}
where $\rho$ is the density which corresponds to $m_0$.
%COMMENT: SO WHAT?
%GOT THIS FAR STEP 2 
\section{Polytropic Gas Spheres and their Stability}

In the present section we solve (\ref{2.5}) through (\ref{2.8}) for 
polytropic gas spheres.  We present four solutions. The first is an 
analytic solution of these equations while the other two utilize
numerical computations. We consider the stability of these solutions.
The stability of the solution is calculated using (\ref{9.3}).

\subsection{Polytropic Sphere with Analytic Solution}

For the present case we start by choosing a functional form for the 
density $\rho(r)$ and then solve (\ref{2.5}) for $m(r)$. Equation
(\ref{2.6}) becomes an algebraic equation for $\lambda(r)$ while 
(\ref{2.7}) is a differential equation 
for $\nu(r)$. Substituting this result in (\ref{7.2}), one 
can compute the isentropy coefficient $A(r)$ (or isentropy index $\alpha(r)$).

The following illustrates this procedure and leads to an analytic
solution for the metric coefficients.

Consider a sphere of radius $R$ (where $0 < R \le \sqrt{2}$) with the 
density function
\begin{equation}
\label{3.2}
\rho(r)= \frac{1}{4}\frac{R^2-r^2}{Br^2}
\end{equation}
where $B$ is the constant in (\ref{2.5}). Using (\ref{2.5}) with the initial
condition $m(0)=0$ we then have for $0 \le r \le R$
\begin{equation}
\label{3.3}
m(r)=\frac{R^2r}{4}-\frac{1}{12}r^3.
\end{equation}
Observe that although $\rho(r)$ is singular at $r=0$ the total mass of 
the sphere is finite.

Using (\ref{2.6}) yields
\begin{equation}
\label{3.31}
\lambda(r)=-ln\left(1-\frac{R^2}{2}+\frac{r^2}{6}\right).
\end{equation}

Substituting (\ref{3.3}) into (\ref{2.12}) we obtain a
general solution for $\nu(r)$ which is valid for $R\ne 1$ and $R\ne \sqrt{2}$. 
\begin{equation}
\label{3.32}
\nu=2\ln(C_1rF(r)^{\omega}+C_2rF(r)^{-\omega})
\end{equation}
where
$$
F(r)=\frac{6-3R^2+\sqrt{6-3R^2}\sqrt{6-3R^2+r^2}}{r},\,\,\,
\omega=\sqrt{\frac{2(R^2-1)}{R^2-2}}.
$$
For R=1 the solution is
\begin{equation}
\label{3.4}
\nu=2\ln\left[r\left(D_1 +D_2 arctanh\sqrt{\frac{3}{3+r^2}}\right)\right].
\end{equation}
At $r=0$ we have $\nu(0)=-\infty$ and the metric is singular at this point.
This reflects the fact that the density function (\ref{3.2}) has 
a singularity at $r=0$ (but the total mass of the sphere is finite). 
We observe that this singularity in $\rho$ at $r=0$ does not correspond 
to any of those classified by Arnold et al \cite{AZ}. This is due to the fact
that none of the solutions presented in \cite{AZ} has a periodic structure.

To determine the constants $D_1$ and $D_2$ we use the fact that at $R=1$ 
the value of $\nu$ should match the classic Schwarzschild exterior solution
$$
e^{\nu(R)}=1-\frac{2M}{R}
$$
and the pressure (see \ref{2.8}) is zero. These conditions lead to the 
following equations:
\begin{equation}
\label{3.41}
\left(D_1+D_2arctanh\frac{\sqrt{3}}{2}\right)^2-\frac{2}{3}=0
\end{equation}
\begin{equation}
\label{3.42}
3D_1+3D_2arctanh\frac{\sqrt{3}}{2}-2\sqrt{3}D_2=0.
\end{equation}
The solution of these equations is
$$
D_1=-\frac{\sqrt{2}}{6}\left( 3~arctanh\frac{\sqrt{3}}{2}-2\sqrt{3}\right),\,\,\,
D_2=\frac{\sqrt{2}}{2}.
$$
Using (\ref{2.8}) we obtain the following expression for the pressure
$$
p=\frac{1}{C}\left\{\frac{D_2\sqrt{3(3+r^2)}}
{3r^2\left(D_1 +D_2 arctanh\sqrt{\frac{3}{3+r^2}}\right)}-
\frac{1}{2}(\frac{1}{r^2}+1)\right\} .
$$
Assuming that $p(r)=A(r)\rho(r)$ we depict $A(r)$ for this 
solution in Fig. \ref{figure 2}.

Note that trying to model this result by a relationship of the form
$p(r)=A\rho^{\alpha(r)}$ leads to discontinuities in the values of $\alpha(r)$.
For $R=\sqrt{2}$ the differential equation for $\nu$ is
\begin{equation}
\label{3.43}
2\frac{d^2\nu}{dr^2}+\left(\frac{d\nu}{dr}\right)^2+\frac{24}{r^4}=0.
\end{equation}
The solution of this equation is
\begin{equation}
\label{3.44}
\nu=-\ln(24)+2\ln \left[r\left(E_1\sin\frac{\sqrt{6}}{r}+E_2\cos\frac{\sqrt{6}}{r}\right)\right]
\end{equation}
and applying the boundary conditions on $\nu$ and the pressure at $r=\sqrt{2}$
we find that
$$
E_1=2\sin(\sqrt{3}),\,\,\,
E_2=2\cos(\sqrt{3}).
$$
If we assume that the relationship between the pressure and the density 
is of the form $p=A\rho^{\alpha(r)}$ then $\alpha(r)$ exhibits several 
local spikes in the range $0 < r <\sqrt{2}$ but is zero otherwise.

The plot for a perturbation  $m_1(r)$ from the initial mass distribution 
$m_0(r)$ in  (\ref{3.3}) is presented in Fig. \ref{Figure 3}. This figure demonstrates
that using (\ref{9.3}) this mass distribution is stable to perturbations of order $m_1(0)=10^{-3}$.

\subsection{Spheres with Oscillatory Density Functions}

Here we discuss several examples of spheres with oscillatory density functions and 
determine the appropriate polytropic index (or coefficient) that describes 
these spheres. We probe also for the stability of these mass configurations
to small perturbations.

\subsubsection{Infinite Sphere with Exponentially Decreasing Density}
Let
\begin{equation}
\label{14.14}
\rho(r)=e^{-r} (D+\cos r) ,\,\,\, 0\le r \le \infty
\end{equation}
where $D=1.1$. The deviation of $D$ from $1$ is needed  to avoid
$\rho=0$ in (\ref{7.60})-(\ref{7.61}). Otherwise these equations 
become singular when $\rho=0$.

It follows from (\ref{2.6}) (with $m(0)=0$) that
\begin{equation}
\label{14.15}
m(r)=B\left\{(2D-\frac{1}{2})-\frac{e^{-r}}{2}
\left[(r^2-1)\cos(r)-(r+1)^2\sin(r)+2D((r+1)^2+1)\right]\right\}
\end{equation}
Observe that although the sphere is assumed to be of infinite radius
the density approaches zero exponentially as $r\rightarrow \infty$
and the total mass of the sphere is finite.

Substituting these expressions in (\ref{7.60}) with $A=B=1$ and $D=1.1 $ and 
solving for $\alpha(r)$ we obtain Fig. \ref{Figure 4} which exhibits a strong decline in 
the value of $\alpha(r)$ as the density decreases exponentially. If we 
substitute $B=1$, $\alpha=1$ and $D=1.1$ in (\ref{7.61} we obtain Fig. $5$  
where $A(r)$ has a steep negative gradient as $\rho(r)\rightarrow 0$. 

The plot for a perturbation $m_1(r)$ from $m_0(r)$ that is given by 
(\ref{14.15}) is presented in Fig. \ref{Figure 6} (using (\ref{9.3})).It shows that 
the mass distribution remains stable to perturbations whose order is 
$m_1(0)=10^{-5}$. 

\subsubsection{Finite Sphere with Ring Structure}

We consider a sphere of radius $\pi$ with density function
\begin{equation}
\label{3.21a}
\rho=\frac{sin^2(kr)}{k^2r^2}.
\end{equation}
>From (\ref{2.5}) with $m(0)=0$ we then have
\begin{equation}
\label{3.22a}
m(r)=\frac{B[2kr-\sin(2kr)]}{4k^3}
\end{equation}
where the total mass $M$ of the sphere is $\frac{B\pi}{2k^2}$.

Fig. \ref{Figure 7} depicts the solution of (\ref{7.60}) for $\alpha(r)$
with $A=1$, $B=1$ and $k=4$. This figure exhibits a steep downward slope 
in the value of $\alpha(r)$ beyond $r=0.8$ due to the decrease in the density. 
Fig. \ref{Figure 8} displays the solution of (\ref{7.61}) for $A(r)$ with $\alpha=1$ and
the same values for $B$ and $k$. The spikes in the values of $A(r)$ in
this figure reflect the thermodynamics processes that are ongoing due 
to the oscillations in the density.

The plot for a perturbation  $m_1(r)$ from $m_0(r)$ given by (\ref{3.22a})
is presented in Fig. \ref{Figure 9} (using the model of (\ref{9.3})). It shows that 
the mass distribution remains stable to perturbations whose order is given 
by $m_1(0)=10^{-5}$.

\subsubsection{Infinite Sphere with Ring Structure}
 
Consider a sphere of infinite radius with the density function
\begin{equation}
\label{3.11}
\rho=\frac{1}{r^2 k^2} e^{-\beta r}(D+\sin(kr)^2)
\end{equation}
where $\beta$, $k$ are constants and $D=0.01$. 

Solving (\ref{2.5}) with the initial condition $m(0)=0$ yields
\begin{eqnarray}
\label{3.12}
&&m(r)=-\frac{B}{2\beta k^2(\beta^2+4k^2)} \\ \notag
&&\left\{e^{-\beta r}[(2D+1)(\beta^2+4k^2)-\beta^2\cos(2kr)+2\beta k\sin(2kr)]
-2D(\beta^2+4k^2)-4k^2\right\}
\end{eqnarray}
Observe that although the sphere is assumed to be of infinite radius
the density approaches zero exponentially as $r\rightarrow \infty$
and the total mass of the sphere is finite.

Fig. \ref{Figure 10} depicts the solution of (\ref{7.60}) for $\alpha(r)$
with $A=1$, $B=1$, $\beta=0.001$ and $k=8$. Similarly Fig. \ref{Figure 11}
displays the solution of (\ref{7.61}) for $A(r)$ with $\alpha=1$ and
the same values for $B$, $\beta$ and $k$.

If we interpret the density function (\ref{3.11}) as one that corresponds
to the density of a primordial gas cloud with ring structure then the 
results shown in Fig. \ref{Figure 10} and Fig. \ref{Figure 11} demonstrate 
that the thermodynamic activity within the cloud is reflected by the 
oscillatory behavior of $A(r)$ and $\alpha(r)$.

As to stability we found that when a polytropic model $p=A(r)\rho$
is used to describe the gas then it is stable only for for perturbations
with $m_1(0) \le 5\times 10^{-9}$. A plot of $m_1(r)$ under this assumption
is presented in Fig. \ref{Figure 12}. However when we assumed that 
$p=A\rho^{\alpha(r)}$ the mass density in the cloud remained stable for 
perturbations satisfying $m_1(0)\le 10^{-2}$ and we obtained Fig. 
\ref{Figure 13}. This demonstrates that in this particular case
the mass distribution (\ref{3.12}) has a much larger basin of stability 
when the gas can be modeled by the relationship $p=A\rho^{\alpha(r)}$.

\section{Conclusions}
In this paper we considered the steady states of a spherical protostar  
or interstellar gas cloud where general relativistic 
considerations are taken into account. In addition we considered 
the gas to be polytropic, thereby removing the (implicit or explicit) 
assumption that it is isothermal. Two polytropic models for the gas
were considered, the first in the form $p=A\rho(r)^{\alpha(r)}$
and the second in the form $p=A(r)\rho(r)^{\alpha}$. 
Under these assumptions we were able to 
derive a single equation for the total mass of the sphere as a function of 
$r$, from whose solution the corresponding metric coefficients 
can be computed in straightforward fashion. Using the TOV equation  
we derived equations for $\alpha(r)$ and $A(r)$. We proved that when
either $\alpha$ or $A$ are constants the mass density of the sphere
cannot be chosen arbitrarily. We derived also an equation for stability 
of these configurations to perturbations in mass density.

Using several idealized models for the density within
primordial gas clouds we were able to compute the appropriate polytropic 
coefficient and index and thus gain new insights about their thermodynamic 
structure. In particular we showed that the mass distribution of a gas cloud
with ring structure can be stable to perturbations. The evolution of this 
ring structure in time (within the framework of General Relativity) will 
be investigated in a subsequent paper.

We conclude then that General Relativity can provide new and deeper
insights about the actual structure of stars and primordial gas clouds
and the emergence of density patterns within these objects.  

To our best knowledge these solutions represent a new and different class 
of interior solutions to the Einstein equations which have not been 
explored in the literature.

\newpage
\begin{figure}[ht]
\centerline{\includegraphics[height=120mm,width=140mm,clip,keepaspectratio]{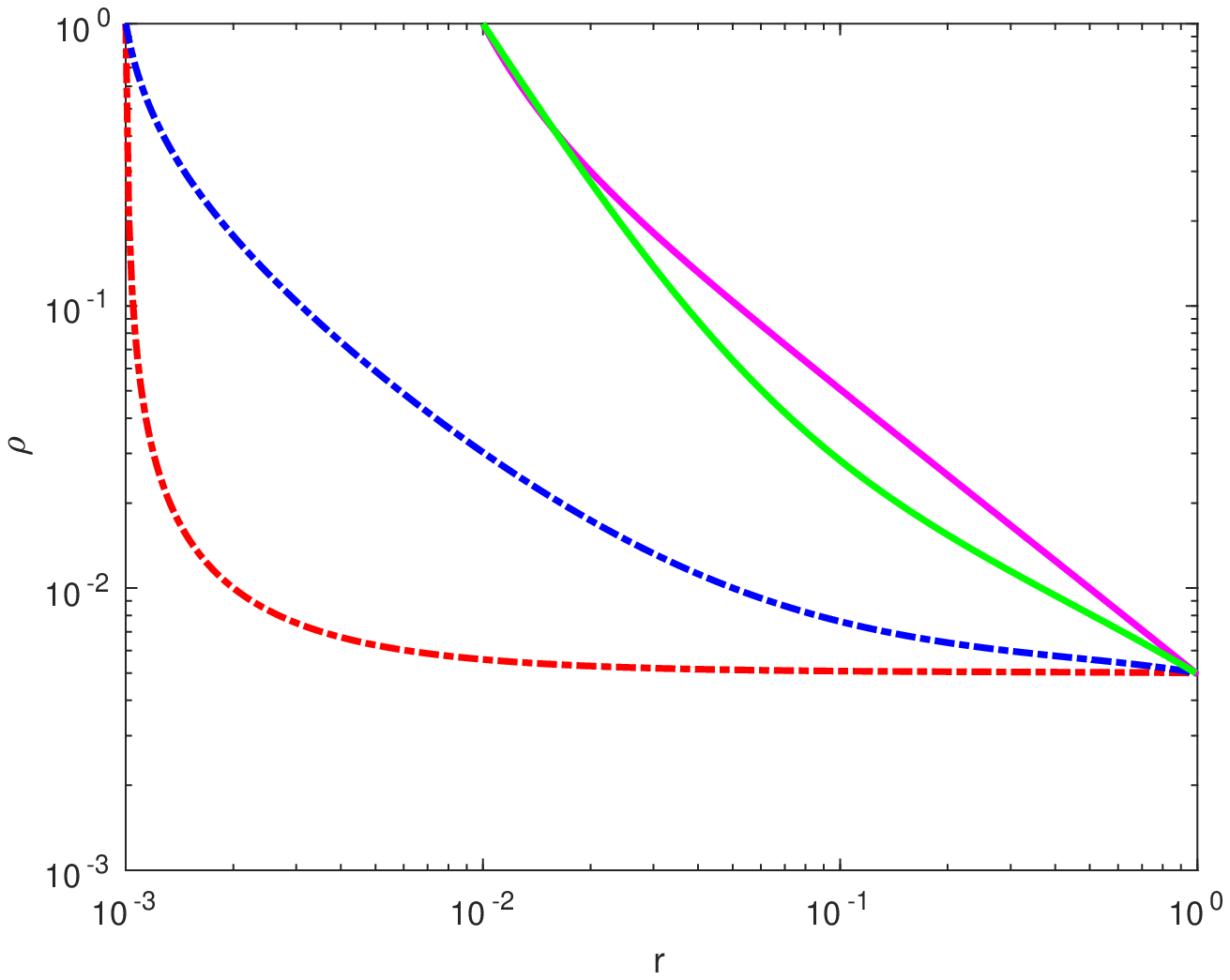}}
\caption{$\rho(r)$ for $A=1$ with $\alpha=1$ and $\alpha=2$ (red and blue 
dashed lines) and for $A=r$ with $\alpha=1$ and $\alpha=2$ (magenta and green solid lines)}
\label{Figure 1}
\end{figure}
\newpage

\begin{figure}[ht]
\centerline{\includegraphics[height=120mm,width=140mm,clip,keepaspectratio]{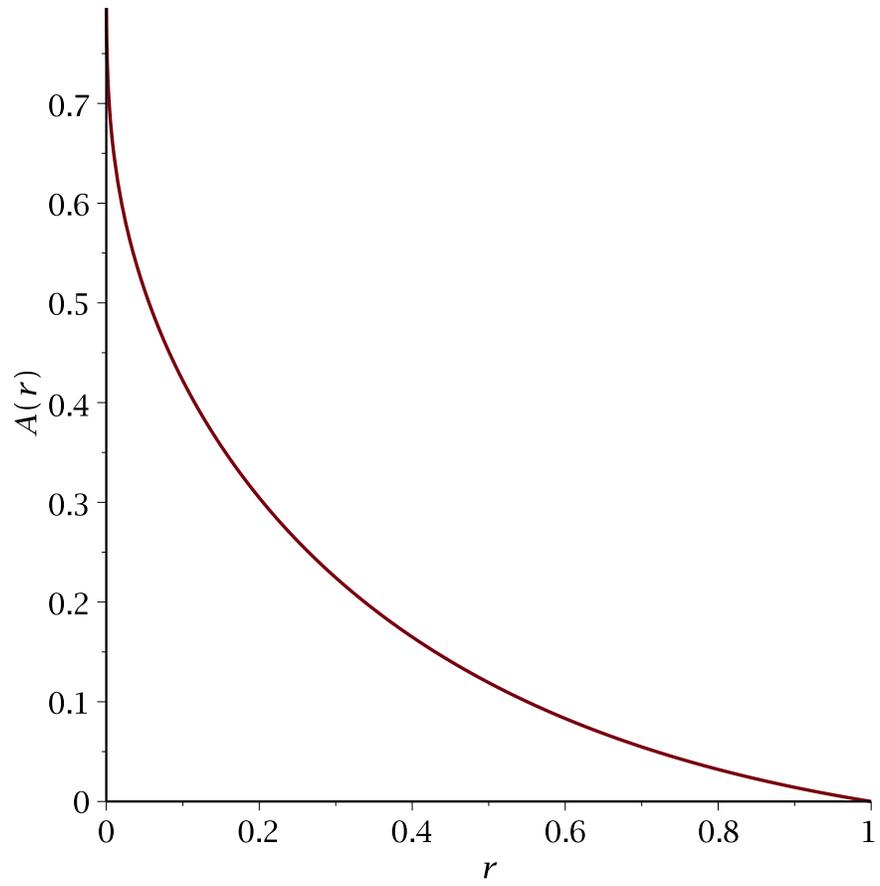}}
\caption{A(r) for the mass distribution (\ref{3.3})}
\label{figure 2}
\end{figure}

\newpage
\begin{figure}[ht]
\centerline{\includegraphics[height=120mm,width=140mm,clip,keepaspectratio]{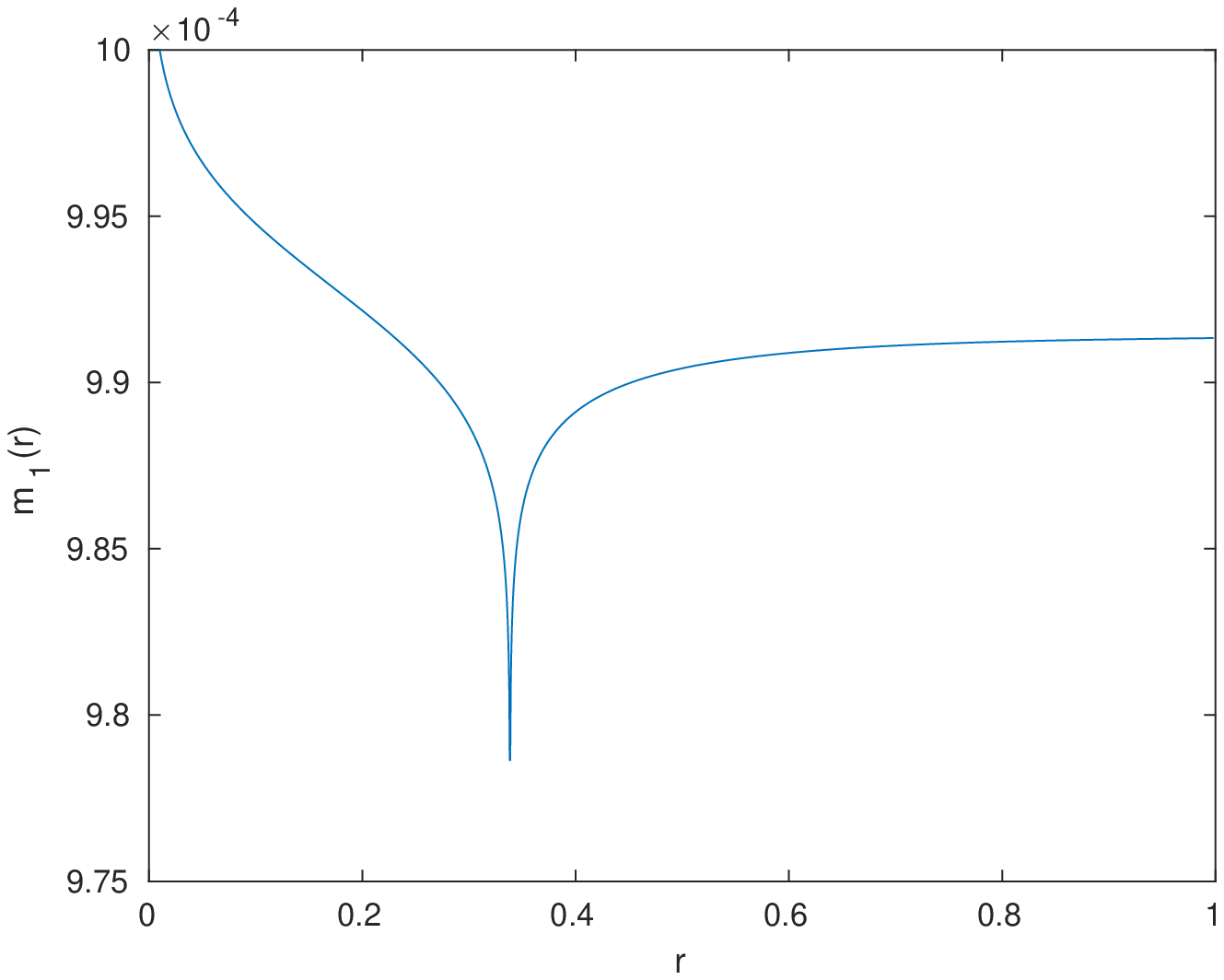}}
\caption{$m_1(r)$ for $m(r)$ in equation (\ref{3.3})}
\label{Figure 3}
\end{figure}

\newpage
\begin{figure}[ht]
\centerline{\includegraphics[height=120mm,width=140mm,clip,keepaspectratio]{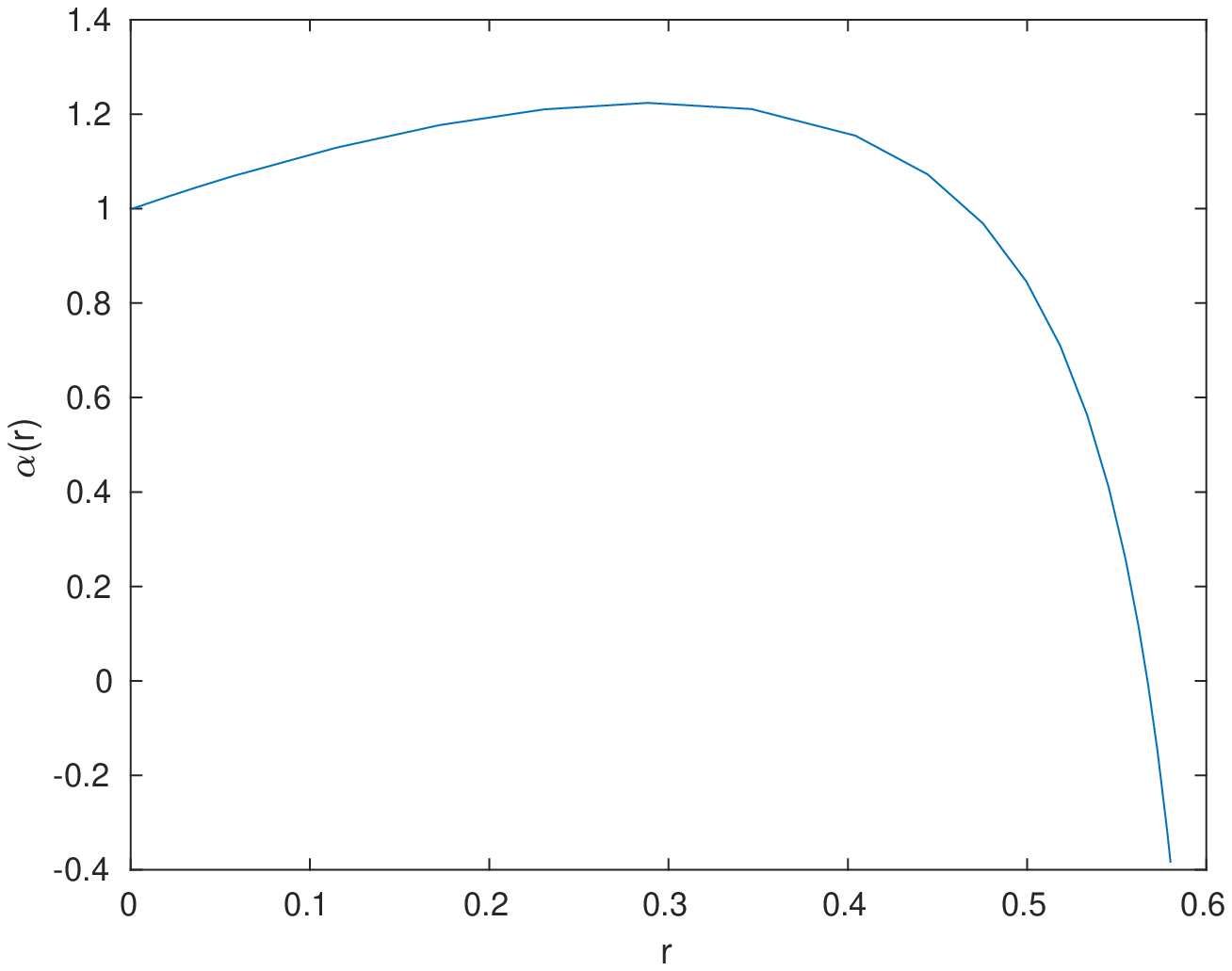}}
\caption{Solution of (\ref{7.60}) for $\alpha(r)$ with $\rho$ in (\ref{14.14}) }
\label{Figure 4}
\end{figure}

\newpage
\begin{figure}[ht]
\centerline{\includegraphics[height=120mm,width=140mm,clip,keepaspectratio]{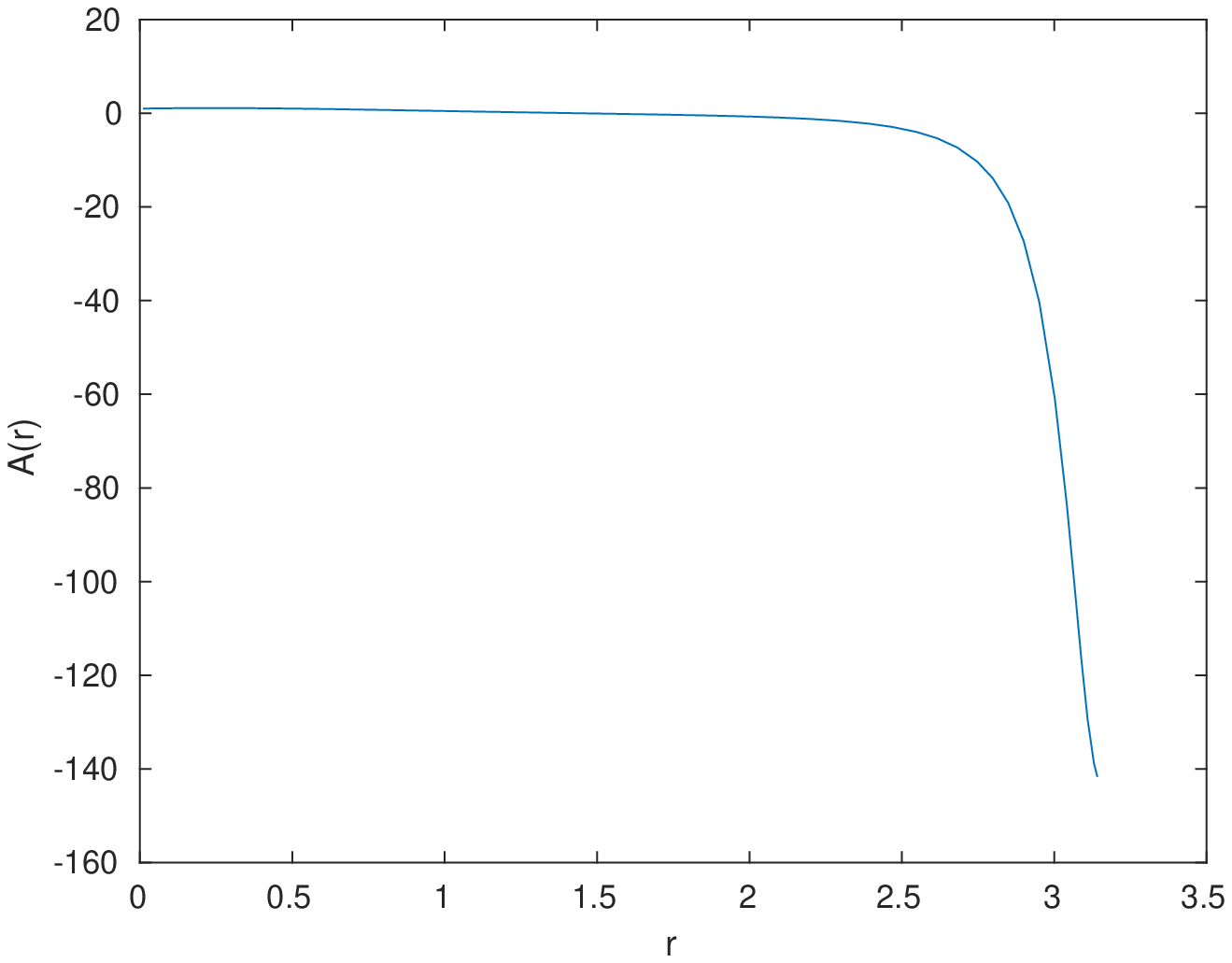}}
\caption{Solution of (\ref{7.61}) for $A(r)$ with $\rho$ in (\ref{14.14}) }
\label{Figure 5}
\end{figure}

\newpage
\begin{figure}[ht]
\centerline{\includegraphics[height=100mm,width=120mm,clip,keepaspectratio]{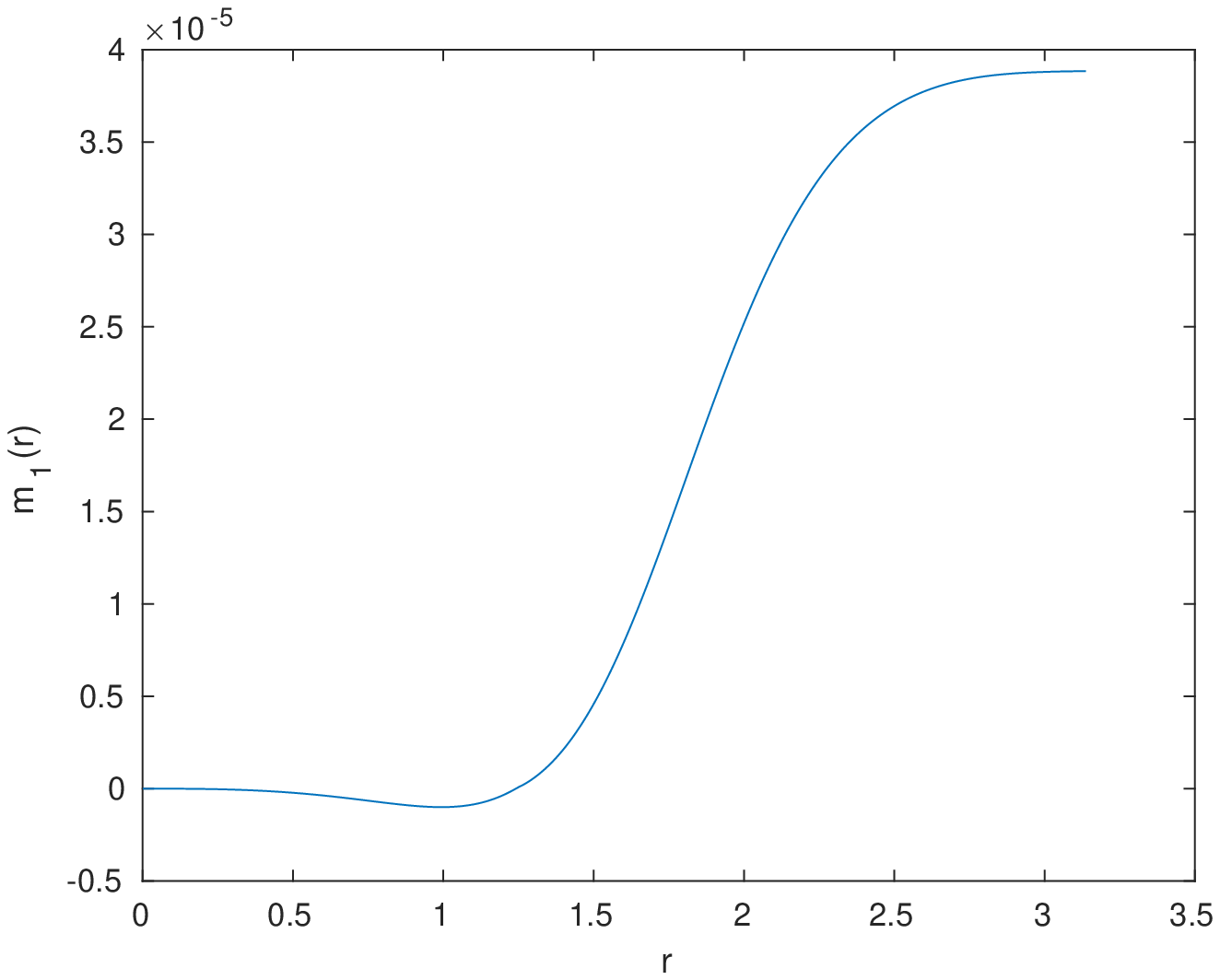}}
\caption{$m_1(r)$ for $m(r)$ in equation (\ref{14.15})}
\label{Figure 6}
\end{figure}

\newpage
\begin{figure}[ht]
\centerline{\includegraphics[height=120mm,width=140mm,clip,keepaspectratio]{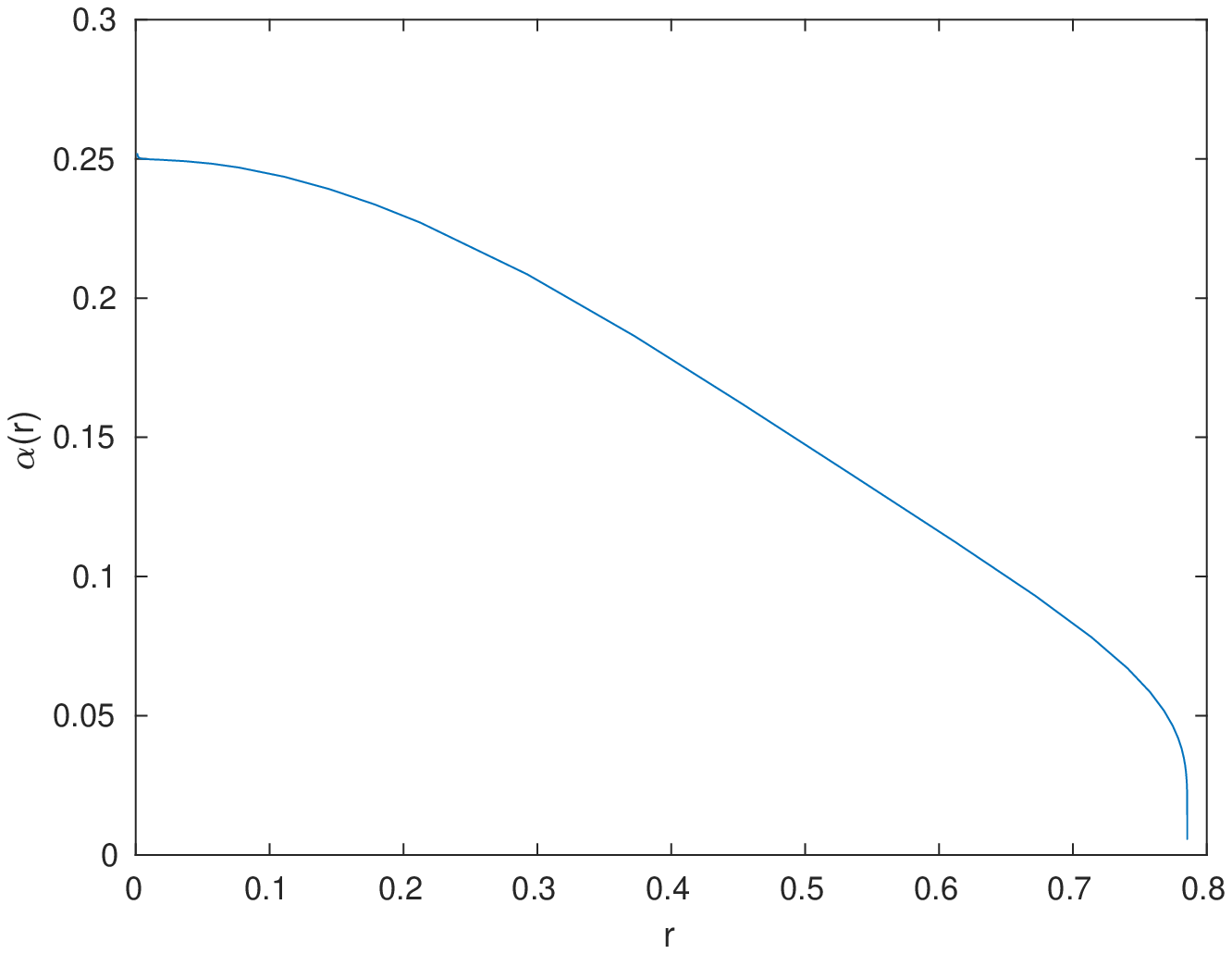}}
\caption{Solution of (\ref{7.60}) for $\alpha(r)$ with $\rho$ in (\ref{3.21a}) }
\label{Figure 7}
\end{figure}

\newpage
\begin{figure}[ht]
\centerline{\includegraphics[height=120mm,width=140mm,clip,keepaspectratio]{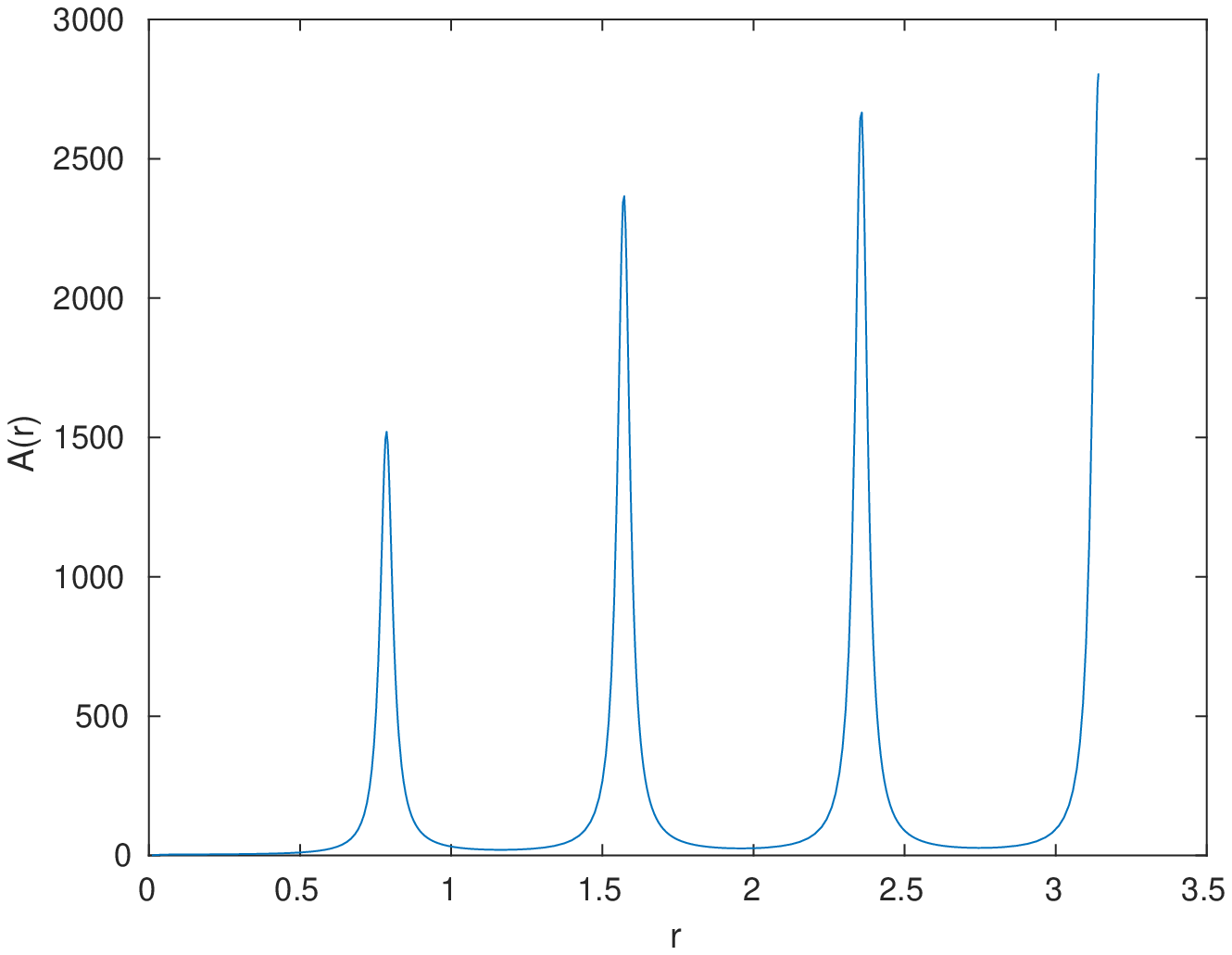}}
\caption{Solution of (\ref{7.61}) for $A(r)$ with $\rho$ in (\ref{3.21a}) }
\label{Figure 8}
\end{figure}

\newpage
\begin{figure}[ht]
\centerline{\includegraphics[height=100mm,width=120mm,clip,keepaspectratio]{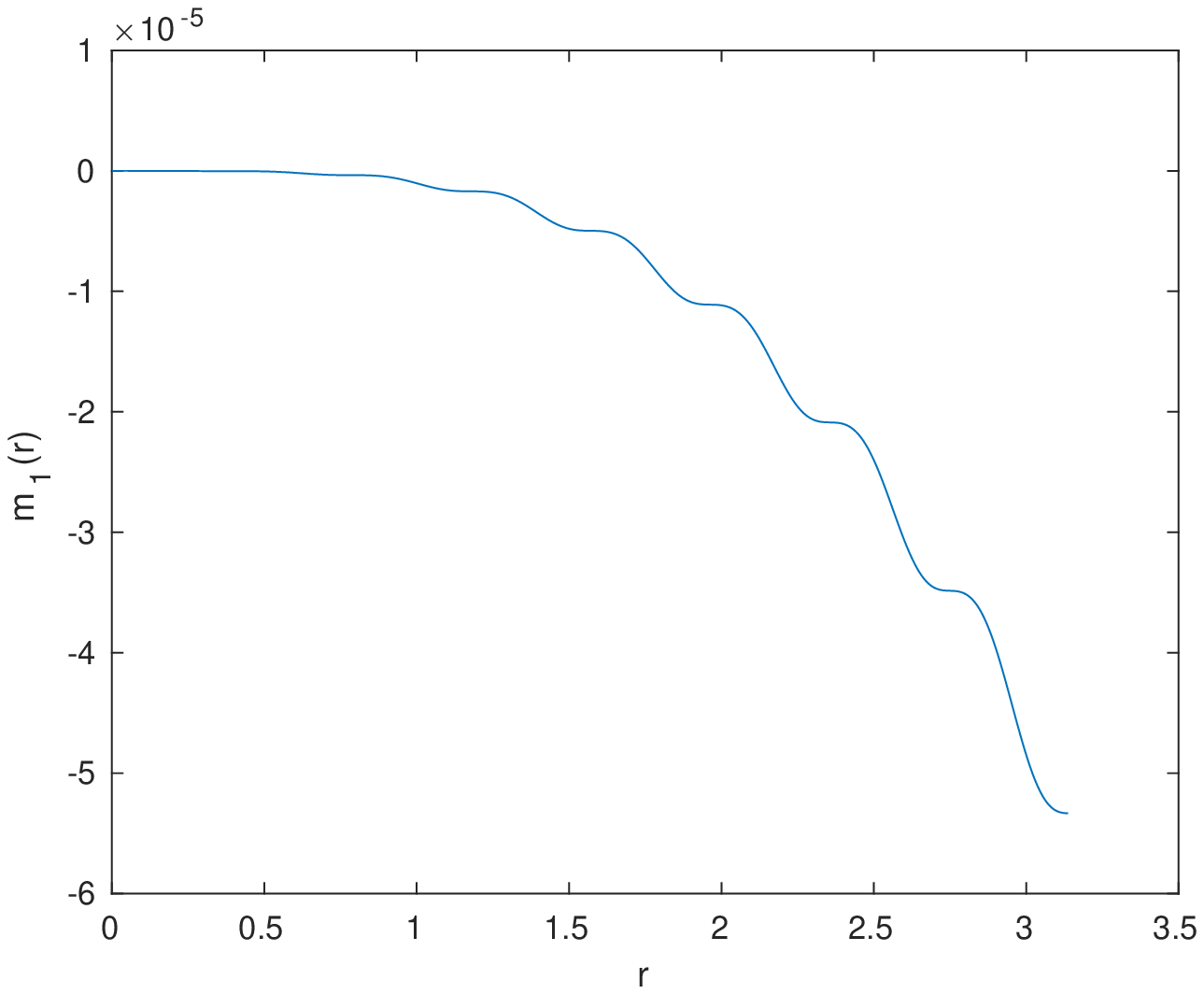}}
\caption{$m_1(r)$ for $m(r)$ in equation (\ref{3.22a})}
\label{Figure 9}
\end{figure}

\newpage
\begin{figure}[ht]
\centerline{\includegraphics[height=120mm,width=140mm,clip,keepaspectratio]{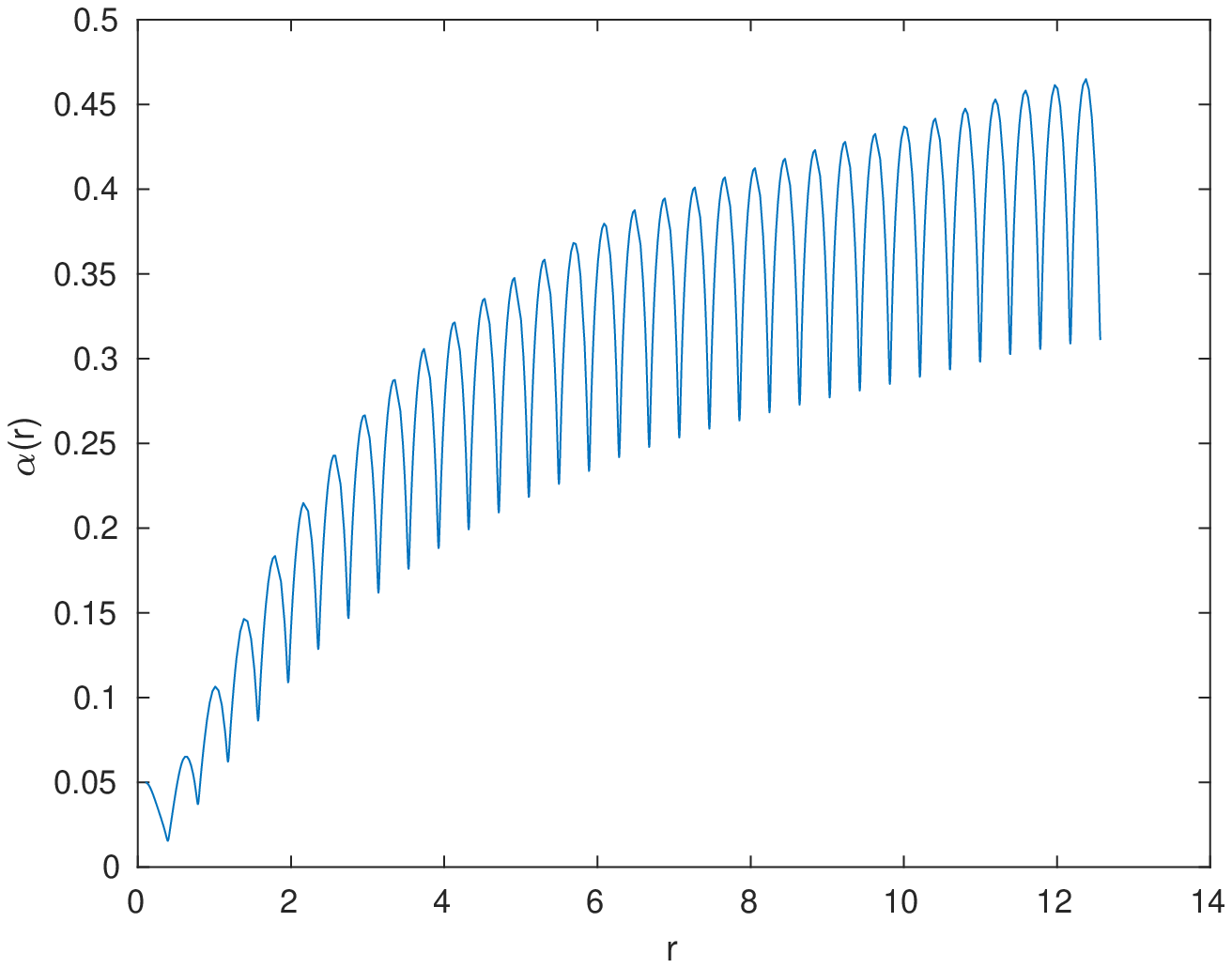}}

\caption{Solution of (\ref{7.60}) for $\alpha(r)$ with $\rho$ in (\ref{3.12}) }
\label{Figure 10}
\end{figure}

\newpage
\begin{figure}[ht]
\centerline{\includegraphics[height=120mm,width=140mm,clip,keepaspectratio]{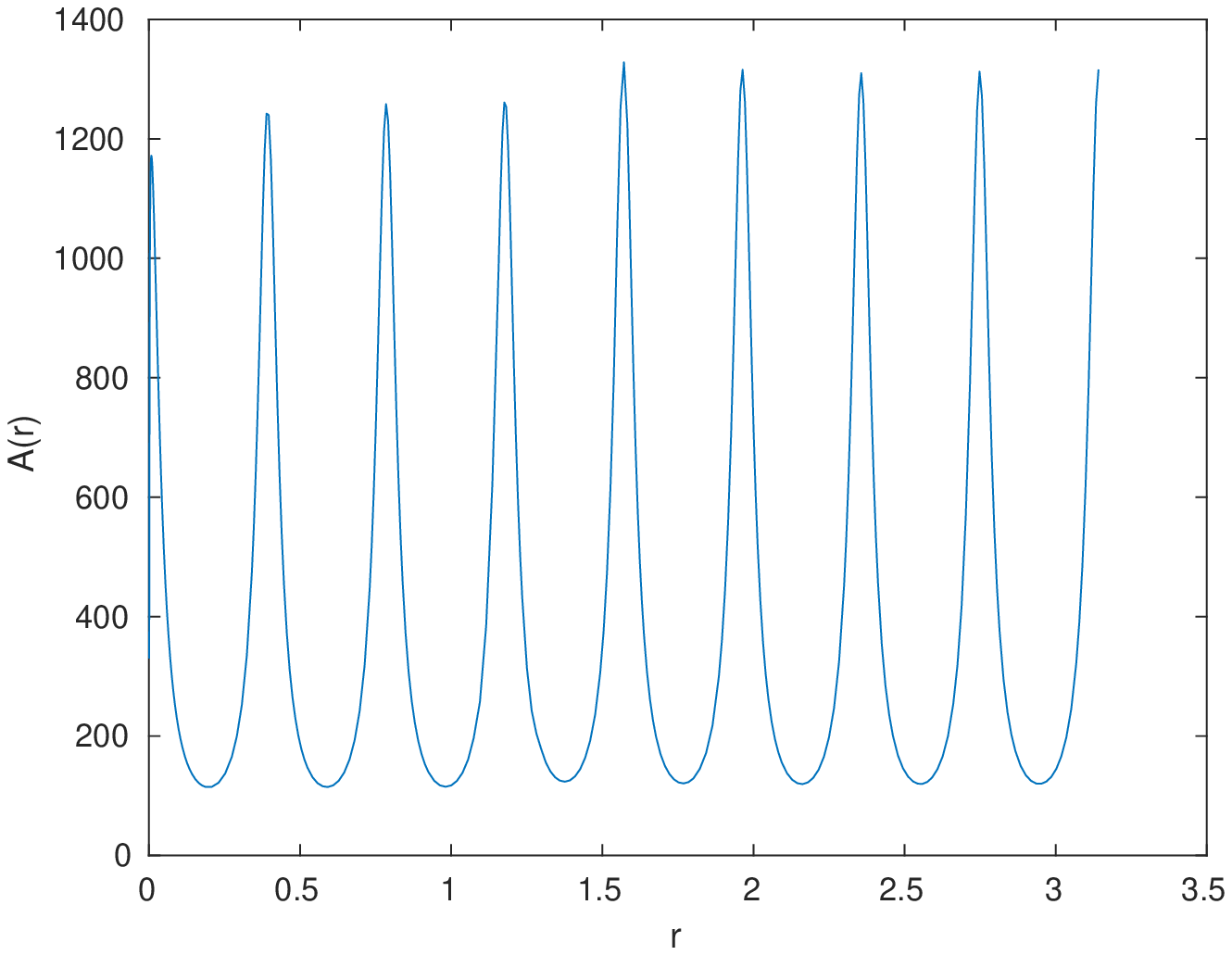}}
\caption{Solution of (\ref{7.61}) for $A(r)$ with $\rho$ in 
(\ref{3.12}) }
\label{Figure 11}
\end{figure}

\newpage
\begin{figure}[ht]
\centerline{\includegraphics[height=120mm,width=140mm,clip,keepaspectratio]{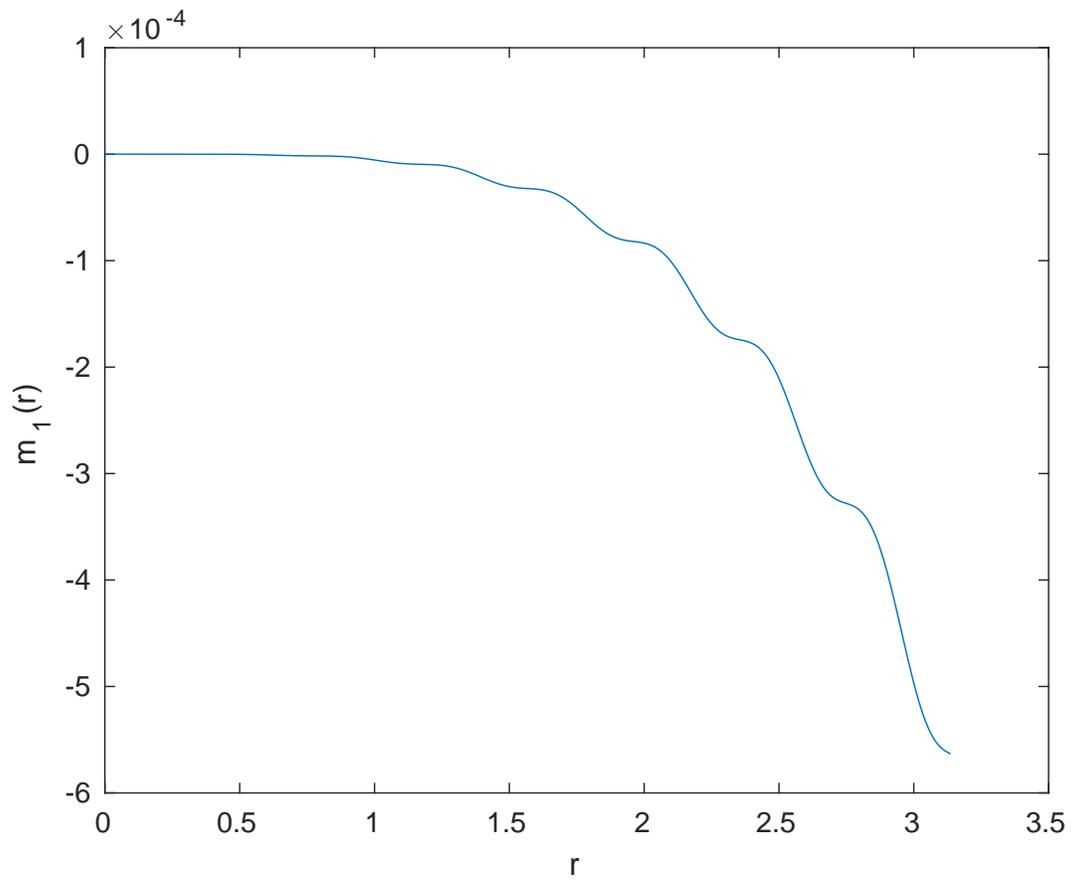}}
\caption{$m_1(r)$ for $m(r)$ in equation (\ref{3.12}) with $p=A(r)\rho$}
\label{Figure 12}
\end{figure}

\newpage
\begin{figure}[ht]
\centerline{\includegraphics[height=120mm,width=140mm,clip,keepaspectratio]{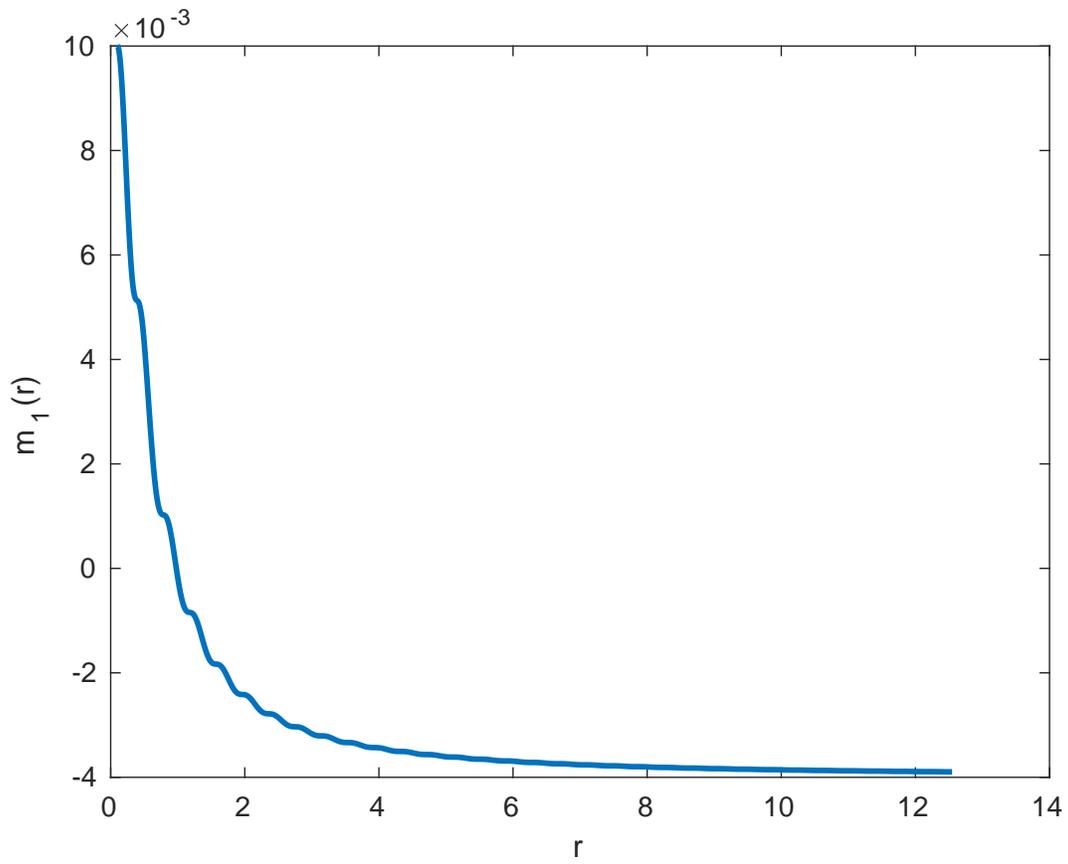}}
\caption{$m_1(r)$ for $m(r)$ in equation (\ref{3.12}) with $p=A\rho^{\alpha(r)}$}
\label{Figure 13}
\end{figure}
\end{document}